\documentclass[12pt]{article}
\usepackage{graphicx}

\newcommand\pubnumber{IFT-UAM/CSIC-22-103}
\newcommand\pubdate{\today}

\newcommand{\soh}{{\textstyle \scriptstyle \frac{1}{2}}}

\newcommand{\smoh}{{\textstyle \scriptstyle - \frac{1}{2}}}

\def\institute{Instituto de F\'\i sica Te\'orica, IFT-UAM/CSIC, \\ c/ Nicolás Cabrera 13-15, 28049 Madrid}
\def\support{\footnote{Work supported by 
 the grants IFT Centro de Excelencia Severo Ochoa CEX2020-001007-S and  PID2019-110058GB-C21, funded by MCIN/AEI/10.13039/501100011033 and by ERDF, and by FCT project CERN/FIS-PAR/0004/2019. }}

\def\Title#1{\begin{center} {\Large #1 } \end{center}}
\def\Author#1{\begin{center}{ \sc #1} \end{center}}
\def\Address#1{\begin{center}{ \it #1} \end{center}}

\newcommand\pubblock{\rightline{\begin{tabular}{l} \pubnumber\\
         \pubdate  \end{tabular}}}
\newenvironment{Abstract}{\begin{quotation}  }{\end{quotation}}
\newenvironment{Presented}{\begin{quotation} \begin{center} 
             PRESENTED AT\end{center}\bigskip 
      \begin{center}\begin{large}}{\end{large}\end{center} \end{quotation}}
\def\Acknowledgements{\bigskip  \bigskip \begin{center} \begin{large}
             \bf ACKNOWLEDGEMENTS \end{large}\end{center}}

\def\beq{\begin{equation}}
\def\eeq#1{\label{#1}\end{equation}}
\def\eeqn{\end{equation}}

\def\beqa{\begin{eqnarray}}
\def\eeqa#1{\label{#1}\end{eqnarray}}
\def\eeqan{\end{eqnarray}}

\let\bar=\overbar

\def\Dslash{\not{\hbox{\kern-4pt $D$}}}
\def\dslash{\not{\hbox{\kern-2pt $\del$}}}

\def\msb{{\bar{\ssstyle M \kern -1pt S}}}

\begin{document}
\begin{titlepage}
\pubblock

\vfill
\Title{Understanding template methods \\[1mm] for top polarisation measurements}
\vfill
\Author{ J. A. Aguilar-Saavedra\support}
\Address{\institute}
\vfill
\begin{Abstract}
Top polarisation measurements at the LHC are often performed using template methods. I discuss the role of quantum interference between polarisation states --- which is mostly overlooked in the literature. Furthermore, I argue
which is the correct definition of the `templates' of definite polarisations, in order to match the experimental measurements with meaningful theoretical predictions.
\end{Abstract}
\vfill
\begin{Presented}
$15^\mathrm{th}$ International Workshop on Top Quark Physics\\
Durham, UK, September 4--9, 2022
\end{Presented}
\vfill
\end{titlepage}
\def\thefootnote{\fnsymbol{footnote}}
\setcounter{footnote}{0}

\section{Introduction}

Before discussing template methods for top quark polarisation measurements, it is worthwhile clarifying what is meant by `polarisation'. The top quark has a mean lifetime around $5 \times 10^{-25}$ s, therefore we can only make observations on its decay products. The processes that involve the production and decay of top quarks have the top quarks as intermediate particles, with amplitudes summing over all polarisation states. However, because the top quark mass is much larger than its width, the narrow-width approximation is accurate and one can decompose the amplitude $\mathcal{M}$ into production ($\mathcal{P}$) and decay factors ($\mathcal{D}$),
\begin{equation}
\mathcal{M} =  \sum_{s = \pm 1/2} \mathcal{P}_s \mathcal{D}_s \,, 
\label{ec:M}
\end{equation}
with $s$ labelling the eigenvalues of the spin operator in some arbitrary $\hat z$ direction, not necessarily the helicity. This expansion follows from the completeness relation for Dirac spinors $\sum_s u(p,s) \bar u(p,s) = \not\!p + m$, in standard notation, with $m$ the fermion mass. Taking the modulus square of (\ref{ec:M}),
\begin{equation}
|\mathcal{M}|^2 = \sum_{s,s' = \pm 1/2} \left( \mathcal{P}_s \mathcal{P}_{s'}^* \right) \mathcal{D}_s \mathcal{D}_{s'} ^* \,.
\end{equation}
The terms between parentheses, summed over different sub-processes, multiplied by phase-space factors, parton distribution functions and properly normalised, give rise to the matrix elements $\rho_{ss'}$ of the spin-density operator describing the spin state in which the top quarks are produced. The polarisation in the $\hat z$ direction, for example, is the difference $\rho_{\soh \soh} - \rho_{\smoh \smoh}$~\cite{Kane:1991bg}, which in turn depends on differences $|\mathcal{P}_\soh |^2 - |\mathcal{P}_\smoh|^2$. The latter are sensitive to new physics contributions in the top quark production. To that end, if we are able to measure the top quark spin density matrix --- which is possible, from angular distributions~\cite{Aguilar-Saavedra:2014eqa,Aguilar-Saavedra:2017wpl} --- we can investigate new physics in the top quark production, which justifies the interest of these  measurements. To be more specific, let us fix a reference system $(x,y,z)$ in the top quark rest frame. Then, the spin state of the top quark is described by the so-called polarisation vector $\vec P = (P_x,P_y,P_z)$, with $|\vec P| \leq 1$, which determines the top spin density matrix
\begin{equation}
\rho = \frac{1}{2} \left(\! \begin{array}{cc}
1 + P_z & P_x - i P_y \\
P_x + i P_y & 1 - P_z
\end{array} \! \right) \,.
\end{equation}
In systems with more than one top quark, additional observables characterise their spin correlation~\cite{Bernreuther:2015yna}.

\section{Template methods}

Template methods can be applied in the same fashion for measurements in single and pair production of top quarks. For simplicity, we will discuss single production, considering top quarks (not anti-quarks). It is quite common to use the charged lepton $\ell = e,\mu$ from the top quark decay $t \to W^+  b \to \ell^+ \nu b$ to measure the top polarisation properties. The normalised charged lepton three-momentum is $\hat p_\ell = (\sin \theta \cos \phi,\sin \theta \sin \phi,\cos \theta)$, in terms of the polar coordinates $(\theta,\phi)$. Integrating the rest of variables, the normalised top production cross section is~\cite{Boudjema:2009fz}
\begin{equation}
\frac{1}{\sigma} \frac{d\sigma}{d\Omega} = \frac{1}{4\pi} \left[ 1 + \alpha_\ell \,  \vec P \cdot \hat p_\ell \right] \,,
\label{ec:dist2D}
\end{equation}
with $d\Omega = d\!\cos \theta d\phi$, and $\alpha_\ell$ the so-called `spin analysing power' constant, $\alpha_\ell^+ = 1$ at the leading order in the standard model (SM). 
If we are interested in the measurement of the diagonal elements of $\rho$, we can integrate over $\phi$ to obtain the well-known distribution
\begin{equation}
\frac{1}{\sigma} \frac{d\sigma}{d\!\cos \theta} = \frac{1}{2} \left[ 1 + \alpha_\ell \,  P_z \cos \theta \right] \,.
\label{ec:dist1D}
\end{equation}
Note that we can write (\ref{ec:dist1D}) as
\begin{equation}
\frac{1}{\sigma} \frac{d\sigma}{d\!\cos \theta}
 = a_+ \left[ \frac{1}{\sigma} \frac{d\sigma}{d\!\cos \theta} \right]_+
+ a_- \left[ \frac{1}{\sigma} \frac{d\sigma}{d\!\cos \theta} \right]_- \,,
\label{ec:t1}
\end{equation}
with $a_+ = (1 + P_z)/2 = \rho_{\soh \soh}$, $a_- = (1 - P_z)/2 = \rho_{\smoh \smoh}$.
The terms between brackets correspond to taking $P_z = 1$ ($+$) and $P_z = -1$ ($-$) in (\ref{ec:dist1D}). The interference between $P_z = 1$ and $P_z = -1$ contributions, which is apparent in (\ref{ec:dist2D}), identically cancels in (\ref{ec:t1}). This
equation shows that the diagonal elements of $\rho$ can be determined by fitting the measured distribution (on the l.h.s.) as a linear combination of two `templates', the normalised distributions between brackets corresponding to $P_z = \pm 1$. The coefficients in this combination are no other than the cross sections for production of top quarks in a $S_z$ eigenstate, divided by the total cross section, $a_\pm = \sigma_\pm / \sigma$. 

Detector resolution and acceptance cuts change this picture~\cite{Aguilar-Saavedra:2021ngj}. In order to derive the template expansion in this case, it is convenient to rewrite (\ref{ec:t1}) as
\begin{equation}
\frac{d\sigma}{d\!\cos \theta} = \frac{d\sigma_+}{d\!\cos \theta} 
+ \frac{d\sigma_-}{d\!\cos \theta} \,.
\label{ec:t2}
\end{equation}
Denoting with bars the quantities after cuts, we can write 
\begin{equation}
\frac{d \bar \sigma}{d\!\cos \theta} = \frac{d \bar \sigma_+}{d\!\cos \theta} 
+ \frac{d\bar \sigma_-}{d\!\cos \theta} + \dots \,,
\label{ec:t2}
\end{equation}
where the dots stand for interference terms that no longer cancel, because the integration is not performed over the full $\phi$ range due to the kinematical cuts. (In addition, the mis-reconstruction of the $\phi$ angle due to detector effects prevents the cancellation of the interference.) This expression can be cast in a form similar to (\ref{ec:t1}) with normalised distributions, by defining `efficiencies' $\varepsilon = \bar \sigma / \sigma$, $\varepsilon_\pm = \bar \sigma_\pm / \sigma_\pm$,
\begin{equation}
\varepsilon \left[ \frac{1}{\bar \sigma} \frac{d \bar \sigma}{d\!\cos \theta} \right]
 = \varepsilon_+ a_+ \left[ \frac{1}{\bar \sigma_+} \frac{d \bar \sigma_+}{d\!\cos \theta} \right] 
+ \varepsilon_- a_- \left[ \frac{1}{\bar \sigma_-} \frac{d\bar \sigma_-}{d\!\cos \theta} \right] + \Delta_{\mathrm int} \,.
\label{ec:t3}
\end{equation}
This derivation highlights two important aspects that have not always been taken into account in current measurements:
\begin{itemize}
\item[(i)] For the template measurement to be sound, the interference term $\Delta_{\mathrm int}$ must be included --- or it has to be shown that it is negligible, compared to other uncertainties.
\item[(ii)] The templates used for the measurement, i.e. the terms between brackets in the r.h.s. of (\ref{ec:t3}), must  correspond to the distributions obtained by projecting to third spin component $\pm 1/2$, obtained for example with spin projectors.
\end{itemize}
We illustrate these two points in the next section.

\section{Lessons to take}

Let us consider single top quark production (not including anti-quarks) and the helicity basis with three orthogonal vectors $(\hat r,\hat n,\hat k)$ defined as
\begin{itemize}
\item K-axis (helicity): $\hat k$ is a normalised vector in the direction of the top quark three-momentum in the centre-of-mass (c.m.) frame.
\item R-axis: $\hat r$ is in the production plane, and is defined by $\hat r = \mathrm{sign}(z_p) (\hat p_p - \cos \theta \; \hat k)/\sin \theta$, with $\hat p_p = (0,0,1)$ the direction of one proton in the laboratory (lab) frame, $\cos \theta = \hat k \cdot \hat p_p$, and $z_p = \vec p_j \cdot \hat p_p$, with $\vec p_j$ the three-momentum of the spectator jet in the lab frame.
\item N-axis: $\hat n = \hat k \times \hat r$ is orthogonal to the production plane.
\end{itemize}
Single top quark production is generated in the four-flavour scheme ($pp \to t \bar b j$) using {\scshape Protos}~\cite{protos}. The templates corresponding to definite polarisations $P = \pm 1$ in the K, R and N axes are obtained by implementing spin projectors at the matrix-element level, generating small samples and subsequently applying the CAR method~\cite{Aguilar-Saavedra:2022kgy} to augment the sample size up to $10^6$ events, the same size of the SM samples. A detector simulation is not performed; instead, a cut on transverse momenta $p_T \geq 30$ GeV of the charged lepton and $b$ quark from the top decay is applied, to distort the distributions with respect to the parton-level ones. The efficiencies for the SM sample and the six templates are
\begin{eqnarray}
\varepsilon_{\mathrm SM} = 0.547 \,, \quad
\varepsilon_{K+} = 0.631 \,, \quad \varepsilon_{K-} = 0.530 \,,  \nonumber \\
\varepsilon_{R+} = 0.603 \,, \quad \varepsilon_{R-} = 0.546 \,, \quad
\varepsilon_{N+} = 0.580 \,, \quad \varepsilon_{N-} = 0.581 \,.
\end{eqnarray}
The interference term $\Delta_{\mathrm int}$ is calculated in the SM as described in Ref.~\cite{Aguilar-Saavedra:2021ngj}: Using (\ref{ec:t3}) with the above efficiencies and the SM predictions for $a_\pm$, the interference contribution is computed by subtracting from a SM sample (on the l.h.s.) the template contributions on the r.h.s., all obtained with samples that are statistically independent from the ones later used for the fit. Of course, the robustness of this calculation in the presence of new physics contributions has to be assessed case by case. In Ref.~\cite{Aguilar-Saavedra:2021ngj}, it was found that for $t \bar t$ production the term $\Delta_{\mathrm int}$ is quite insensitive to anomalous $gtt$ chromo-magnetic couplings that change the $t \bar t$ polarisation.

The effect of taking into account (or not) $\Delta_{\mathrm int}$ in the template fit, already at the partonic level, is illustrated in Table~\ref{tab:int}. While the contribution is not important for the K and N axes, omitting $\Delta_{\mathrm int}$ causes a serious bias in the polarisation measurement for the R axis. In a realistic setup the effect of $\Delta_{\mathrm int}$ is much more prononced. Not only the integration does not cover the full $\phi$ range due to cuts, but also the $\phi$ angle is not well determined due to the mis-reconstruction of the $\hat z$ axis. Consequently, the inclusion of the interference term is crucial also for the K and N axes~\cite{Aguilar-Saavedra:2021ngj}.

\begin{table}[h]
\begin{center}
\begin{tabular}{lccccccccc}
& \multicolumn{3}{c}{K} & \multicolumn{3}{c}{R} & \multicolumn{3}{c}{N} \\
                                           &$a_+$ & $a_-$ & $P$ & $a_+$ & $a_-$ & $P$ & $a_+$ & $a_-$ & $P$ \\
True                                    & 0.159  & 0.841 & -0.682 & 0.615  & 0.385 & 0.230 & 0.501 & 0.499 & 0.002 \\
no $\Delta_{\mathrm int}$   & 0.160 & 0.840 & -0.680 & 0.580  & 0.420 & 0.160  & 0.500  & 0.500 & 0.000 \\
with $\Delta_{\mathrm int}$ & 0.160 & 0.840 & -0.680 & 0.616  & 0.384 & 0.232 & 0.500  & 0.500 & 0.000
\end{tabular}
\caption{`True' values of the polarisation coefficients $a_\pm$ and polarisations $P = a_+ - a_-$ extracted from the Monte Carlo sample without cuts, and values extracted from the template fits. Monte Carlo statistical uncertainties on $a_\pm$ are of the order of $10^{-3}$.}
\label{tab:int}
\end{center}
\end{table}

The question of how to generate the polarised samples that have to be used for the fit can be addressed with a similar exercise. Rather than performing kinematical cuts on top decay products, we perform a cut on the transverse momentum of the second $\bar b$ quark, $p_T^{\bar b} > 30$ GeV, which keeps the top decay distributions unaltered and, therefore, the interference terms vanishing. We restrict ourselves to the polarisation measurement in the K axis. The efficiencies of this kinematical cut for the SM sample and the two templates are
\begin{eqnarray}
\varepsilon_{\mathrm SM} = 0.292 \,, \quad
\varepsilon_{K+} = 0.524 \,, \quad \varepsilon_{K-} = 0.245 \,.
\label{ec:eff2}
\end{eqnarray}
The fit to (\ref{ec:t3}) yields $a_+ \varepsilon_+ / \varepsilon_{\mathrm SM} = 0.286$, $a_- \varepsilon_- / \varepsilon_{\mathrm SM} = 0.714$. Plugging the efficiencies (\ref{ec:eff2}) yields $a_+ = 0.160$, $a_- = 0.840$, $P = -0.680$, in excellent agreement with the true values in Table~\ref{tab:int}. Therefore, in obtaining the correct $a_\pm$ it is not only essential to have the correct normalised angular distributions in the templates, but also to have the correct production kinematics that corresponds to polarised tops, which determines the different efficiencies in (\ref{ec:eff2}). The correct kinematics is obtained by using spin projectors at the matrix-element level.

To further illustrate the difference, one can use the CAR method to generate alternative templates with the (correct) top decay angular distributions corresponding to $P = \pm 1$, but the (incorrect) SM production kinematics. The efficiency of the cut on $p_T^{\bar b}$ for those templates is the same as in the SM, $\varepsilon_{K \pm}^\prime = \varepsilon_{\mathrm SM}$ by construction. Plugging these efficiencies into the values obtained from the fit, one obtains quite incorrect results $a_+ = 0.286$, $a_- = 0.714$, $P = -0.428$.

We conclude by pointing out that, despite the two caveats here highlighted, the application of template methods is quite robust in the `vicinity' of the SM, as the terms $\Delta_{\mathrm int}$ can be well estimated by the SM ones, and the production kinematics is similar. Should data significantly depart from the SM --- which unfortunately is not the case --- the template methods would have to be further refined.

\Acknowledgements
I am grateful to the organisers of the TOP 2022 conference for the invitation, which prompted further studies on the topic, and to M.L. Mangano and J. Alcaraz for very useful discussions. I also thank the Galileo Galilei Institute For Theoretical Physics for hospitality during the finalisation of this work.

\end{document}